\documentstyle[prl,aps]{revtex}
\begin{document}
\draft
\tighten
\twocolumn[\hsize\textwidth\columnwidth\hsize\csname 
@twocolumnfalse\endcsname
\title{Interplay between Zeeman Coupling and Swap Action
in Spin-based Quantum Computer Models: Error Correction
in Inhomogeneous Magnetic Fields}
\author{Xuedong Hu, Rogerio de Sousa, and S. Das Sarma}
\address{
Department of Physics, University of Maryland, College Park,
MD 20742-4111
}
\date{\today}
\maketitle
\begin{abstract}
We consider theoretically the interplay between Zeeman coupling 
and exchange-induced swap action in spin-based quantum dot
quantum computer models in the presence of inhomogeneous magnetic 
fields, which are invariably present in real systems.  We
estimate quantitatively swap errors caused by the inhomogeneous 
field, establishing that error correction would, in principle,
be possible in the presence of non-uniform magnetic fields
in realistic structures.
\end{abstract}

\vskip2pc]
\narrowtext

In recent years quantum computing has attracted widespread 
attention.  The computers based on the 
principles of quantum mechanics, such as quantum parallelism
and entanglement, promise to deliver results much faster than
classical computers in certain tasks such as factoring
and searching \cite{Reviews}.  
The impetus for constructing real quantum
computer architectures arose from the seminal 
results \cite{error_correction} establishing that quantum error 
correction is theoretically possible and therefore decoherence 
is not an insurmountable barrier as was assumed earlier.  
There have been many proposals 
for quantum computer (QC) architectures based on
various physical two-level systems, such as those using
laser-cooled trapped ions \cite{Zoller,iontrap},
photons or atoms trapped in cavities \cite{Sleator,cavityQED},
nuclear spins in bulk solutions \cite{NMR} and crystal 
lattices \cite{Fumiko},
spins of electrons trapped in quantum dots \cite{LD,Imam} or
donors \cite{Vrijen}, spins of donor nuclei in Si \cite{BKane},
superconducting devices \cite{Schon}, and others.  
The minimal requirements for a QC architecture are the
existence of fundamental quantum bits (qubit) and the ability
to carry out single- and two-qubit operations such as swap and 
controlled-NOT, as well as suitable hardwares for reading 
(input) and writing (output or measurement) operations.
Among the many obstacles facing the successful demonstration 
of nontrivial quantum computation in specific QC hardwares,
the most daunting are the problem of quantum decoherence
and the difficulty in achieving precise control over the
various unitary operations necessary for quantum computation.
In this Letter we consider the theoretical issue of 
controlling the swap operation in the proposed solid state
QC architecture involving quantum dot spin entanglement,
with externally-controlled electrostatic gates and magnetic
fields providing the unitary operation.

Initial work on QC hardwares concentrated on atomic/molecular 
systems partly because well-defined single-qubit operations are 
comparatively easier to control in atomic systems such as 
trapped ions \cite{Zoller,iontrap}.  
Successful QC architectures will eventually
require many qubits working in parallel, which would be difficult 
to achieve in atomic systems.  Proposed solid state QC 
architectures are, in principle, scalable to many qubits,
although no one has yet successfully demonstrated a controlled
single-qubit operation in solid state QC systems.  This is
a curious dichotomy in the current QC hardware research---the
architectures with demonstrated single-qubit operations are
difficult to scale up while the presumably scalable 
solid state architectures have not yet been able to demonstrate
single-qubit control.  It is thus crucial to explore the
challenges facing coherent control of qubits in solid state
structures, particularly the issue of possible error
corrections in realistic systems.
Among various microscopic degrees of freedom that have been
considered for the role of qubits in solid state QC architectures, 
spins of electrons
or nuclei are natural candidates because of their well-defined
Hilbert spaces and their relatively long decoherence time compared
to the orbital degrees of freedom.  In the proposed spin-based
QC architectures,
exchange interaction plays a fundamental role of establishing
two-qubit entanglement \cite{Fumiko,LD,Vrijen,BKane}, while Zeeman
coupling to an external magnetic field provides
various single-qubit operations.  Generally Zeeman coupling
is treated separately from the exchange 
interaction \cite{BLD,HD} because the
latter originates from the Coulomb interaction 
(due to Pauli principle) while the former
is purely a spin effect, and the two interaction terms in the
Hamiltonian commute with each other if the Zeeman term is 
homogeneous.  However, solid state heterostructures are intrinsically
inhomogeneous and magnetic imperfections or impurities are likely 
to be present, leading to inhomogeneous stray magnetic fields.  
Furthermore, parallel pulse schemes, in which exchange interaction
and inhomogeneous Zeeman coupling are present simultaneously, have been 
proposed to expedite operations of spin-based QCs \cite{parallel}.
Therefore, it is necessary to explore the interplay between
exchange interaction and Zeeman coupling
in relation to spin-based QC architectures, as we do here.

In this Letter we focus on the effects of Zeeman coupling
when the external magnetic field is spatially inhomogeneous.  
Zeeman coupling
is generally neglected in these studies \cite{BLD,HD} based 
partly on
the assumption that the effects are small and do not lead to 
any qualitative changes.  While it is true in uniform
fields, this assumption may not hold in an inhomogeneous
magnetic field.  The key motivations for our study are the
observations that spatially inhomogeneous fields are 
intrinsic to spin-based solid state QC architectures 
and that
for parallel schemes the magnetic fields at the locations 
of the two spins are generally different \cite{parallel}, 
which means the total field must be inhomogeneous.  
If only the single-spin evolution is involved, effects of 
local magnetic field can be corrected in such single-qubit
operations by techniques such 
as spin echoes \cite{Slichter}.  However, if two-spin
entangled evolution is also involved, it is not clear
whether errors caused by inhomogeneous 
fields can still be eliminated, {\it i.e.} whether such 
errors lie within the current QC error correction constraints.

As an 
example we study how Zeeman coupling may affect the
proposed operations of spin-based quantum dot quantum computers
(QDQC).  Our results, however, are quite general and can be
applied to other spin-based models with finite magnetic fields.
In particular, we study the 
effect of a finite inhomogeneous magnetic field on swap actions.  
For most of the spin-based schemes, swap action $U_{sw}$, 
in which two spins exchange their states, is one of the most basic
operations.  It is used to construct conditional phase shifts (CPS),
$U_{CPS} = 
e^{i\frac{\pi}{4} \sigma_{1z}}
e^{-i\frac{\pi}{4} \sigma_{2z}}
U_{sw}^{\frac{1}{2}} e^{i\frac{\pi}{2} \sigma_{1z}}
U_{sw}^{\frac{1}{2}}$
(with $\sigma$ being the Pauli matrices for spins),
which can then be converted to controlled-NOT (CNOT) 
easily \cite{LD}.  In addition, 
swaps are used to move spin states around \cite{LD,BKane} 
so that an arbitrary pair of spins can be brought into controlled 
entanglement, which is essential to quantum computation \cite{Reviews}.
It is therefore important to investigate how Zeeman splitting
will affect the swap gate and what are the consequences if
these effects are non-trivial.  

In a single envelope function approach \cite{HD}, the Hamiltonian
for two electrons trapped in a lateral double-quantum-dot is
\begin{eqnarray}
H & = & \sum_{i}^{2} \left[ \frac{1}{2m^*} \left( {\bf p} 
+ \frac{e}{c}{\bf A}({\bf r}_i) \right)^2 + V({\bf r}_i)\right. 
\nonumber \\
& & \left. + g^* \mu_B {\bf B}({\bf r}_i)
\cdot {\bf S}_i \right] + \frac{e^2}{\epsilon r_{12}} \,.
\label{eq:Hamiltonian}
\end{eqnarray}
Notice that here the magnetic field has a spatial dependence.
The Hamiltonian can be expanded in a basis
of two-spin eigenstates (spin singlet and triplet states).
If the magnetic field is uniform, the Zeeman coupling
depends only on the total spin along the field direction and
commutes with the total spin operator ${\bf S}^2$, so that
singlet and triplet states remain eigenstates of
the two-electron system.  On the other hand, if the magnetic 
field is inhomogeneous, it destroys the symmetry between the 
two spins, and the singlet state and one of the triplet
states (the one with $S_{\bf n}=0$, where ${\bf n}$ is the field 
direction) couple.  For example the coupling between the lowest 
singlet and triplet states (in the Heitler-London approximation)
is
$\langle T| V_Z |S\rangle = g\mu_B/2 \times \left\{ 
\langle L | B_z({\bf r}) | L \rangle
- \langle R | B_z({\bf r}) | R \rangle 
+ \langle L | B_z({\bf r}) | R \rangle 
\langle R | L \rangle\right.$\break 
$\left.- \langle R | B_z({\bf r}) | L \rangle 
\langle L | R \rangle \right\}$,
where the singlet state is
$|S\rangle = \left[ |L (1)\rangle 
|R (2)\rangle + |L (2)\rangle |R (2)\rangle
\right] (|\!\uparrow\downarrow\rangle 
- |\!\downarrow\uparrow\rangle)/2$, the triplet state is
$|T\rangle = \left[|L (1)\rangle 
|R (2)\rangle - |L (2)\rangle |R (2)\rangle
\right] (|\!\uparrow\downarrow\rangle$\break 
$+|\!\downarrow\uparrow\rangle)/2$, and the Zeeman coupling is
$V_Z = g \mu_B \left[ B_z({\bf r}_1) S_{1z} 
+ B_z({\bf r}_2) S_{2z} \right]$.
Here we have assumed that the external field is along the z
direction.  $|L \rangle$ and $|R \rangle$ are the 
left and right quantum dot ground states.
Therefore, the inhomogeneous magnetic field couples the singlet 
and the $S_z=0$ triplet states 
so that two of the eigenstates of 
the system are no longer the eigenstates of the total spin.
When the overlap between the two ground state 
wavefunctions is small, the main contribution to this 
coupling comes from the average field
difference between the two quantum dots.
Below we explore the consequences of this loss of symmetry.

When the overlap of the electronic wavefunctions is small, we
can assume that the electron orbital degrees of freedom are 
frozen, so that an effective spin Hamiltonian quite faithfully
describes the two-spin system \cite{BLD,HD}:
\begin{eqnarray}
H_s & = & J {\bf S}_1 \cdot {\bf S}_2 + \gamma_1 S_{1z} 
+ \gamma_2 S_{2z} \,,
\label{eq:spin-Hamiltonian} 
\end{eqnarray}
where $\gamma_1 = g \mu_B \langle L | B_z({\bf r}) | L 
\rangle$ and 
$\gamma_2 = g \mu_B \langle R | B_z({\bf r}) | R \rangle$
are local Zeeman couplings due to applied or stray magnetic fields.
Here we have implicitly assumed, based on the small interdot
wavefunction overlap, that the two spins are distinguishable, 
with spin 1 on the left dot and spin 2 on the right dot.
We have also assumed that the field is 
entirely along the z direction.  Whether such a choice is
reasonable will be discussed later.

Hamiltonian (\ref{eq:spin-Hamiltonian})
can be expressed in the basis of four two-spin
states $|\!\uparrow\uparrow\rangle$, 
$|\!\uparrow\downarrow\rangle$,
$|\!\downarrow\uparrow\rangle$, 
and $|\!\downarrow\downarrow\rangle$,
and its eigenstates can be easily obtained.
The two polarized states are decoupled from the other two,
which are mixtures of singlet and $S_z=0$ triplet states:
$|\psi_1\rangle = |\!\uparrow\uparrow\rangle$,
$|\psi_2\rangle = |\!\downarrow\downarrow\rangle$,
$|\psi_3\rangle = c_1 |\!\uparrow\downarrow\rangle
+ c_2 |\!\downarrow\uparrow\rangle$, and
$|\psi_4\rangle = c_2 |\!\uparrow\downarrow\rangle
- c_1 |\!\downarrow\uparrow\rangle$.  Here the coefficients
$c_1$ and $c_2$ satisfy the relations
$c_2/c_1 = \sqrt{1+(\delta/2J)^2} - \delta/2J$ and
$c_1^2+c_2^2 = 1$, where $\delta=\gamma_1-\gamma_2$
represents the field inhomogeneity.
The energies of the latter two states are also shifted from
those of singlet and triplet states (with $\tilde{\delta}$
being the shift):  
$E_1 = J + \Delta$,
$E_2 = J - \Delta$,
$E_3 = -J + \sqrt{4J^2 + \delta^2} \ = \ J + \tilde{\delta}$, and
$E_4 = -J - \sqrt{4J^2 + \delta^2} \ = \ -3J - \tilde{\delta}$,
where $\Delta=\gamma_1+\gamma_2$ represents the average magnetic 
field.  An important question 
here is whether these mixtures and shifts will 
cause any error in quantum computation in the schemes
based on the exchange interaction.
After all, the swap action in these models depends on the 
perfect phase matching in the evolution of singlet and triplet
states, as we will show below.  Since swap operation is an
essential component of a spin-based QDQC and several other 
architectures,
we need to precisely quantify the effects of mixtures in singlet
and triplet states on the swap action.

To determine whether swap is affected, we explore
whether a product state of spins 1 and 2 will evolve into
a product state (pure states for both spins) again.  Our strategy
here is to calculate the Schmidt number of the two-spin state.
We will evolve our two-spin state, calculate its density
matrix, trace out the second spin so that the density matrix of 
the first spin is left.  We can then look for the eigenvalues 
of this density matrix. If at sometime it has only one non-vanishing
eigenvalue, the state of the first spin is pure.  We can then find out
whether this pure state corresponds to a swapped state.

Our initial state is a product state given by
\begin{equation}
|\phi(0)\rangle=(\alpha_1 |\!\uparrow\rangle + \alpha_2 
|\!\downarrow\rangle) (\beta_1 |\!\uparrow\rangle + \beta_2 
|\!\downarrow\rangle) \,.
\end{equation}
If the two electrons are located in two well-separated quantum
dots in the beginning, the above product state does not violate
the antisymmetry requirement of a two-fermion state.  This state
can be expanded in the basis of the eigenstates of Hamiltonian
(\ref{eq:spin-Hamiltonian}).  It then evolves under
Hamiltonian (\ref{eq:spin-Hamiltonian}).  The two-spin state at 
time $t$ takes the form
$|\phi(t)\rangle = \alpha_1 \beta_1 e^{-iE_1 t/\hbar}
|\!\uparrow \uparrow\rangle 
+ \alpha_2 \beta_2 e^{-iE_2 t/\hbar} 
|\!\downarrow \downarrow\rangle 
+ (\alpha_1 \beta_2 c_1 + \alpha_2 \beta_1 c_2) e^{-iE_3 t/\hbar}
|\!\uparrow \downarrow\rangle
+ (\alpha_1 \beta_2 c_2 - \alpha_2 \beta_1 c_1) e^{-iE_4 t/\hbar}
|\!\downarrow \uparrow\rangle$.
The corresponding density matrix for the first spin can then be
calculated straightforwardly:
$\rho_1 = {\rm Tr}_2\{\rho_{12}\} = {\rm Tr}_2\{|\phi(t)\rangle
\langle\phi(t)|\}$.
The matrix elements of $\rho_1$ are
all functions of the constants $\alpha_1$, 
$\alpha_2$, $\beta_1$, $\beta_2$, $c_1$, $c_2$, field 
inhomogeneity $\delta$, exchange coupling $J$, and
time $t$ ($\rho_{1\uparrow\downarrow}$ is also a function of 
the average field $\Delta$).
The eigenvalue equation for $\rho_1$ is
\begin{equation}
\lambda^2 - (\rho_{1\uparrow\uparrow}+\rho_{1\downarrow\downarrow})
\lambda + (\rho_{1\uparrow\uparrow}\rho_{1\downarrow\downarrow}
-|\rho_{1\uparrow\downarrow}|^2) = 0\,.
\end{equation}
To have a pure state for the first spin, which means that only
one eigenvalue of the density matrix $\rho_1$ is non-vanishing,
the last term on the left-hand-side of the above equation has to 
vanish:
\begin{equation}
\rho_{1\uparrow\uparrow}\rho_{1\downarrow\downarrow}
-|\rho_{1\uparrow\downarrow}|^2 = 0 \,.
\label{eq:criterion}
\end{equation}

The explicit expression for Eq.~(\ref{eq:criterion}) is quite 
complicated, so let us first look at the simple uniform field 
situation, when $\delta=0$ and $c_1=c_2=1/\sqrt{2}$. 
Equation~(\ref{eq:criterion}) then becomes
$|\alpha_1\beta_1\alpha_2\beta_2 -\frac{1}{4}[(\alpha_1 \beta_2 
+ \alpha_2 \beta_1)+(\alpha_1 \beta_2 - \alpha_2 \beta_1) e^{i\theta}]
[(\alpha_1 \beta_2 + \alpha_2 
\beta_1)-(\alpha_1 \beta_2 - \alpha_2 \beta_1) e^{i\theta}]|^2 = 0$,
where $\theta=4Jt$.
The solutions here are $e^{i\theta}=\pm 1$.  When 
$e^{i\theta}=1$, $|\phi_1(t)\rangle=\alpha_1 
|\!\uparrow\rangle + \alpha_2 e^{i\Delta t}
|\!\downarrow\rangle$, the state of the first spin returns to its
initial state with a phase shift between the two coefficients.
When $e^{i\theta}=-1$, $|\phi_1(t)\rangle=\beta_1 e^{i\Delta t}
|\!\uparrow\rangle 
+ \beta_2 |\!\downarrow\rangle$, the swap is achieved with the 
exception of an additional phase that can be corrected easily 
with a single-spin operation.
For example, a pulse sequence can be constructed based on the 
phase-shifted swap $U_{psw}$ to produce a CNOT gate
\begin{eqnarray}
U_{CNOT} & = & e^{i\frac{\pi}{4} \sigma_{2y}} \ 
e^{i(\frac{\pi}{4} + \frac{\tilde{\delta} t}{2}) \sigma_{1z}} \ 
e^{-i(\frac{\pi}{4} - \frac{\tilde{\delta} t}{2}) \sigma_{2z}}
\nonumber \\
& & \times U_{psw}^{\frac{1}{2}} \ e^{i\frac{\pi}{2} \sigma_{1z}}
\ U_{psw}^{\frac{1}{2}} \ e^{-i\frac{\pi}{4} \sigma_{2y}} \,.
\end{eqnarray}
Physically, a uniform field means that the Zeeman coupling couples
to the total spin (including both electron spins),
so that the Zeeman term commutes with the exchange term in the 
Hamiltonian (\ref{eq:spin-Hamiltonian}), and therefore does not change
the eigenstates.  The shifts in the energy levels of the polarized
states cause additional phase shift, but can be corrected by applying
an opposite magnetic field with the same pulse shape, magnitude,
and length.  In summary, an external uniform magnetic field does
not qualitatively change the proposed QC algorithm.  Logistically
it makes the QC operation more difficult because of the 
necessary correction pulses.  

If the magnetic field is inhomogeneous, $\delta \neq 0$. 
Let us look at a simple situation when the initial
state is $|\phi(0)\rangle=|\!\uparrow \downarrow \rangle$, 
{\it i.e.} $\alpha_1=\beta_2=1$ and $\alpha_2=\beta_1=0$.  
Eq.~(\ref{eq:criterion}) now takes the simple form of
\begin{equation}
|c_1 c_2 (1-e^{i\theta})(c_1^2+c_2^2 e^{i\theta})|^2=0 \,,
\label{eq:criterion_ud}
\end{equation}
where $\theta=(4J+2\tilde{\delta})\,t$.
When $c_1 \neq c_2$, as is the case when $\delta \neq 0$,
the only solution of this equation is $e^{i\theta}=1$, which 
corresponds to the ``return'' operation with an additional phase: 
$|\phi_1(t)\rangle = \alpha_1 |\!\uparrow\rangle 
+ \alpha_2 e^{i\Delta t}|\!\downarrow\rangle$.  
The condition for
an exact swap does not exist anymore.  
To find the best approximation to
a swap (so that the state of first spin is as close as possible
to being ``down'' in the current special case), 
we calculate the minima of the expression in
Eq.~(\ref{eq:criterion_ud}) and find that $e^{i\theta}=-1$
does still produce a minimum in $|(1-e^{i\theta})
(c_1^2+c_2^2 e^{i\theta})|^2$ (which is, however, no longer zero).
When this condition is satisfied, the density matrix of the
first spin is
\begin{equation}
\left. \rho_1 \right|_{e^{i\theta}=-1}
=\frac{1}{1+x^2}|\!\downarrow\rangle \langle \downarrow\!|
+ \frac{x^2}{1+x^2}|\!\uparrow\rangle \langle \uparrow\!| \,,
\label{eq:swap_error}
\end{equation}
where $x=\delta/(2J)$.  Therefore, the state of the first spin 
cannot be simultaneously pure and the same as the initial state
of the second spin.  The state of the first spin
will remain mixed (when it is 
close to a swap) with the state of the second spin.  
According to Eq.~(\ref{eq:swap_error}), there would 
be an error of the magnitude $x^2$ if we perform such a ``swap'' 
operation, which needs to be corrected.  

For GaAs-based QDQC, the Zeeman splitting is $g\mu_B 
\approx 2.55
\times 10^{-2}$ meV/Tesla,  
while a single Bohr magneton 
produces a field of about 5 Gauss at 1 nm distance and 
almost nothing at a distance of 100 nm in GaAs.  This difference
in the magnetic field leads to a $x^2$ in the order of $10^{-6}$,
which is right within the capability of currently available error 
correction schemes.  Since the field
involved in $\delta$ is the average value over an entire quantum 
dot, local magnetic field inhomogeneity caused by 
impurities should not cause intractable errors.  

Mathematically, 
inhomogeneity in the magnetic field means that a
part of the Hamiltonian (\ref{eq:spin-Hamiltonian}) does not commute
with the exchange term $J {\bf S}_1 \cdot {\bf S}_2$, thus the  
eigenstates will change, leading to errors in swap, which will 
cause errors in CNOT in the proposed sequential 
scheme \cite{LD}.  
In the parallel pulse scheme \cite{parallel},
conditional phase shifts can be produced directly
from the Hamiltonian (\ref{eq:spin-Hamiltonian}) (with precisely
controlled field inhomogeneity), 
circumventing the swap actions.  However, stray fields,
which may be invariably present in realistic solid state QDQC
architectures due to magnetic impurities/imperfections, can
still produce errors which need to be corrected
(even in solely exchange-based models which have no applied
magnetic field \cite{DPD}).
Furthermore, the impossibility of exact swap does make the transfer of
spins through swap an error-prone process in the presence of an
inhomogeneous field.  

Finally, 
inhomogeneity in the external magnetic field is 
not only a possible error source in swap and other two-qubit
operations, but also an 
important factor in the single-qubit operations
determined solely by the Zeeman coupling.  Here the main
concern is the precise definition of the direction of the 
quantization axis for the electron spins.  For example, if
spin-up and -down along z direction correspond to the two
states of our qubit, then the z axis is our quantization axis which 
provides a reference frame for all the single-qubit operations. 
In our calculations the magnetic field is purely along 
the z direction.  The only spatial variation is in its 
magnitude.  However, according to 
Maxwell equations, in a steady state, $\nabla \times {\bf B}=0$.
Thus, if $\partial B_z /\partial x \neq 0$, $\partial B_x /
\partial z = \partial B_z /\partial x \neq 0$.  
If we have a perfect
two dimensional QD system lying entirely in the x-y plane, 
we can assume that $B_x$ happens to 
vanish in the plane of the quantum dots.  However,
using the example of GaAs quantum well QD's, the thickness of the
well is generally around $5-10$ nm, which characterizes the
typical z-width of the QD's.  
Therefore, to have a
vanishing Zeeman coupling along x direction, the $B_x$ field
has to vanish along the middle plane of the quantum well
and be an odd function in the z (growth) direction, which is
a quite stringent constraint.  
If we choose
z-direction as our quantizing axis, a finite $B_x$ field,  
which may be unavoidable, would tend to flip the spins.  
For a 1 Gauss $B_x$ field, this flipping rate is about
0.65 MHz, corresponding to a flipping time of 1.5 $\mu$s.  
Considering that the gates should be operated as slowly as 
possible to satisfy the adiabatic condition, this is a
constraint one has to take into consideration.  
In fact, ``decoherence'' induced by a fluctuating inhomogeneous
transverse field may turn out to be an important extrinsic 
decoherence channel for QDQC operations.

This work is supported by ARDA, LPS, and US-ONR.

\end{document}